\documentclass[11pt,a4paper]{article}
\usepackage{jheppub}
\usepackage{color}
\usepackage{subfigure}

\newcommand{\be}{\begin{equation}}
\newcommand{\bea}{\begin{eqnarray}}
\newcommand{\eea}{\end{eqnarray}}
\newcommand{\ba}{\begin{array}}
\newcommand{\ea}{\end{array}}
\newcommand{\ee}{\end{equation}}

\def\bse{\begin{subequations}}
\def\ese{\end{subequations}}

\title{Holographic Josephson Junction in 3+1 dimensions}

\author{Yong-Qiang Wang \footnote{Corresponding author.},}
\author{Yu-Xiao Liu}
\author{and Zhen-Hua Zhao}

\affiliation{Institute of Theoretical Physics, Lanzhou University,
Lanzhou 730000, People¡¯s Republic of China}
\emailAdd{yqwang@lzu.edu.cn, liuyx@lzu.edu.cn, zhaozhh02@gmail.com}

\abstract{
In arXiv:1101.3326[hep-th], a (2+1)-dimensional holographic Josephson junction was constructed, and it was shown that the DC Josephson current is proportional to the sine of the phase difference across the junction. In this paper, we extend this study to a holographic description for the (3+1)-dimensional holographic DC Josephson junction. By solving numerically the
coupled differential equations, we also obtain the familiar characteristics of Josephson junctions.
}


\begin{document}

\maketitle

\section{Introduction}

The study of the AdS/CFT correspondence and its applications have been the subject of great activity
since the original conjecture was made by Maldacena \cite{Maldacena:1997re}.  The AdS/CFT correspondence is a powerful tool to study the strong coupled field theory. In the last few years, the application of the AdS/CFT correspondence to the  Superconducting/Superfluid phase has been an interesting subject, which states that  the Einstein-Maxwell-scalar model provides a holographically dual description of a superconductor~\cite{Gubser:2008px,Hartnoll:2008vx,Hartnoll:2008kx}.  For reviews on
holographic superconductors, see~\cite{Hartnoll:2009sz,Herzog:2009xv,Horowitz:2010gk}.

Recently, in Ref.~\cite{Horowitz:2011dz}, the authors proposed a gravity model which provides a holographically dual description
of a 2+1 dimensional Josephson Junction. By choosing spatial dependence $\mu$ and non-vanishing constant $J$, which correspond to chemical potential and superfluid currents on the AdS boundary \cite{Basu:2008st,Herzog:2008he,Arean:2010xd,Sonner:2010yx}, respectively, the authors have solved numerically  the
coupled, nonlinear partial differential equations and reproduced the well-known result that the DC Josephson current is proportional to the sine of the phase difference across the junction. Meanwhile, they also studied dependence of the maximum current on the temperature and size of the junction, which matches precisely with the results for condensed matter physics.

In this paper, we would like to extend the
work \cite{Horowitz:2011dz} to a holographic description for the (3+1)-dimensional holographic DC Josephson junction.  We will study the 5D Einstein-Maxwell-scalar model  and  solve numerically  the
coupled, nonlinear partial differential equations of motion,
and look for how to construct a holographic Josephson junctions by choosing the boundary conditions.

The paper is organized as follows. In Sec. \ref{sec2}, we
review  5D Einstein-Maxwell-scalar theories  and  set up a gravity dual of a (3+1)-dimensional SNS Josephson junction.
In Sec. \ref{sec3}, we show numerical results of the EOMs
and study the characteristics of the (3+1)-dimensional holographic Josephson junctions. The last section is devoted to conclusion.

 \section{The model}\label{sec2}







In this section, we study the dynamics of a complex scalar field coupled to a U(1) vector potential in  (4+1)-dimensional AdS spacetimes.
The action of gravity can be written as
\be
S=\int d^5x \sqrt{-g}(R+\frac{12}{L^2}) \;,
\ee
where $L$ is the curvature radius of asymptotic AdS Spacetimes.
We are interested in  the solution of  the planar Schwarzschild black hole with the form:
\be\label{matrix}
ds^2=-f(r)dt^2+\frac{dr^2}{f(r)}+r^2(dx^2+dy^2+dz^2)\;,\qquad f(r)=\frac{r^2}{L^2}-\frac{M}{r^2}\;,
\ee
where $M$ is the mass of the black hole. The black hole has the event horizon $r_{H}$ at
\begin{equation}
r_{H}=M^{\frac{1}{4}}L^{\frac{1}{2}}.
\end{equation}
 The Hawking  temperature of the black hole reads as
\be
T =
\frac{1}{4 \pi } \frac{df}{dr}\bigg|_{r=r_H}=\frac{M^{1/4}}{\pi L^{3/2}}.
\ee
In this background, we
now consider a Maxwell field and a charged complex scalar field, with
the action
\begin{eqnarray}
S=\int d^5x\sqrt{-g}\left[
-\frac{1}{4}F^{\mu\nu}F_{\mu\nu}-|\nabla\psi - iA\psi|^2
-m^2|\psi|^2
\right] ,
\end{eqnarray}
where $F_{\mu\nu}=\partial_\mu A_\nu-\partial_\nu A_\mu$.
The equations of motion include the scalar equation
\be \label{phi}
- \left(\nabla_a - i  A_a \right) \left(\nabla^a - i  A^a\right)\psi + m^2 \psi = 0 \,,
\ee
and Maxwell's equations
\be\label{max}
\nabla^a F_{ab} = i  \left[\psi^* (\nabla_b - i  A_b)\psi - \psi (\nabla_b + i  A_b)\psi^* \right]
\,.
\ee
Taking a static ansatz
\be  \label{ansz}
\psi=|\psi|e^{i\varphi}\;,\qquad  A=A_t\;dt+A_r\;dr+A_x\;dx\;,
\ee
where $|\psi|$, $\varphi$, $A_t$, $A_r$, and $A_x$ are real functions of $r$ and $x$, one can
introduce the gauge-invariant field $M=A- d\varphi$, which is the same definition as that in \cite{Horowitz:2011dz}.

With the black hole background (\ref{matrix}) and the above ansatz (\ref{ansz}), the equation of scalar field (\ref{phi}) can be  written as the real part:
\begin{equation}
\partial_{r}^2|\psi|+\frac{1}{r^2f}\partial_{x}^2|\psi|+\left(\frac{f'}{f}+\frac{3}{r}\right)\partial_r|\psi|+\frac{1}{f}\left(\frac{M_t^2}{f}-fM_r^2-\frac{M_x^2}{r^2}-m^2\right)|\psi|=0\;,\label{EO1}\\
\end{equation}
and the imaginary part:
\begin{equation}
\partial_{r}M_r+\frac{1}{r^2f}\partial_xM_x+\frac{2}{|\psi|}\left(M_r\partial_r|\psi|+\frac{M_x}{r^2f}\partial_x|\psi|\right)+\left(\frac{f'}{f}+\frac{3}{r}\right)M_r=0\;.\label{EO2}
\end{equation}
The imaginary part of the scalar equation can also come from the conservation of the source in Maxwell's equations.
Meanwhile, Maxwell's equations~(\ref{max}) can be written as
\addtocounter{equation}{1}
\begin{align}
\partial_{r}^2M_t+\frac{1}{r^2f}\partial_{x}^2M_t+\frac{3}{r}\partial_r M_t-\frac{2|\psi|^2}{f}M_t &=0\;,\label{Mt}\\
\partial_{x}^2M_r-\partial_{r}\partial_xM_x-2r^2|\psi|^2M_r&=0\;,\label{Mr}\\
\partial_{r}^2M_x-\partial_{r}\partial_xM_r+(\frac{f'}{f}+\frac{1}{r})\left(\partial_r M_x-\partial_x M_r\right)-\frac{2|\psi|^2}{f}M_x&=0\;,\label{Mx}
\end{align}
where a prime denotes derivative with respect to $r$.
Because Eqs.~(\ref{EO1})-(\ref{Mx}) are coupled nonlinear equations, one can not solve these equations
analytically. However, it is straightforward to solve them numerically.

In the following part of this paper, we choose the mass of the scalar field to be  $m^2=-3$, which is above the Breitenl\"ohner-Freedman bound~\cite{Breitenlohner:1982bm}. In order to solve these coupled equations, first, we need to impose regularity at the horizon and the boundary on the radial coordinate.
At the horizon $r=r_H,~M_t(r_H)=0$; At the Ads boundary, the scalar field takes the asymptotic form
\be
|\psi|=\frac{\psi^{(1)}(x)}{r}+\frac{\psi^{(2)}(x)}{r^3}
+\mathcal{O}\left(\frac{1}{r^4}\right)\;,
\ee
the asymptotic behavior of the Maxwell fields  are
\begin{align}
M_t&=\mu(x)-\frac{\rho(x)}{r^2}+\mathcal{O}\left(\frac{1}{r^3}\right)\;,\\
M_r&=\mathcal{O}\left(\frac{1}{r^3}\right)\;,\\
M_x&=\nu(x)+\frac{J}{r^2}+\mathcal{O}\left(\frac{1}{r^3}\right)\; \label{defnu},
\end{align}
here, $\mu$, $\rho$, $\nu$ and $J$ are interpreted as the chemical potential, charge density, superfluid velocity and current, respectively \cite{Horowitz:2008bn,Arean:2010zw}. $J$ is a constant.
For $|\psi|$, both $\psi^{(1)}$ and $\psi^{(2)}$ are normalizable, one can impose the condition either $\psi^{(1)}$ or $\psi^{(2)}$ vanishes. For simplicity, we will consider the case $\psi^{(1)}=0$ and interpret $\langle\mathcal{O}\rangle=\psi^{(2)}$  as the condensate.  Here, we introduce the same phase difference  $\gamma = \Delta \varphi -\int A_x$ as the case in \cite{Horowitz:2011dz}, which is the gauge invariant and can be rewritten as
\be
\gamma=- \int_{-\infty}^\infty dx \;[\nu(x) -\nu(\pm\infty)]\;.
\label{gamma}
\ee

Second, on the spatial coordinate $x$,
we impose the Dirichlet-like boundary condition on $M_r$ and  Neumann-like boundary condition on $|\psi|$, $M_t$ and $M_x$ at $x=0$.
At $x=\pm\infty$, the function is $x$-independent. Thus the boundary conditions of the coupled equations (\ref{EO1})-(\ref{Mx}) are determined by $J$ and $\mu$.

According to the study of paper \cite{Horowitz:2011dz}, we can also introduce the critical temperature $T_c$ of the junction , which is proportional to $\mu_{\infty}=\mu(\infty)=\mu(-\infty)$:
\be\label{T_c}
T_c=\frac{1}{\pi}\frac{\mu(\infty)}{\mu_c}\;,
\ee
where $\mu_c\approx 4.16$.  Furthermore, the effective critical temperature inside the gap is
\be\label{T_0}
T_0=\frac{1}{\pi}\frac{\mu(0)}{\mu_c}\;.
\ee
In order to describe an SNS Josephen junction, we choose the  below  $\mu(x)$  which is the same as that in \cite{Horowitz:2011dz}:
\be\label{profile}
\mu(x)=\mu_\infty\left\{1-\frac{1-\epsilon}{2\tanh(\frac{L}{2\sigma})}\left[\tanh\left(\frac{x+\tfrac{L}{2}}{\sigma}\right)-\tanh\left(\frac{x-\tfrac{L}{2}}{\sigma}\right)\right]\right\}\;,
\ee
where $\mu_\infty$ is the chemical potential at $x=\pm\infty$, and $L$ is the width of our junction. The quantities $\sigma$ and $\epsilon$ control the steepness and depth of our profile, respectively.

\section{Numerical results}\label{sec3}

\begin{figure*}
\begin{center}
 \subfigure[$M_t$]{
  \includegraphics[width=0.4\textwidth]{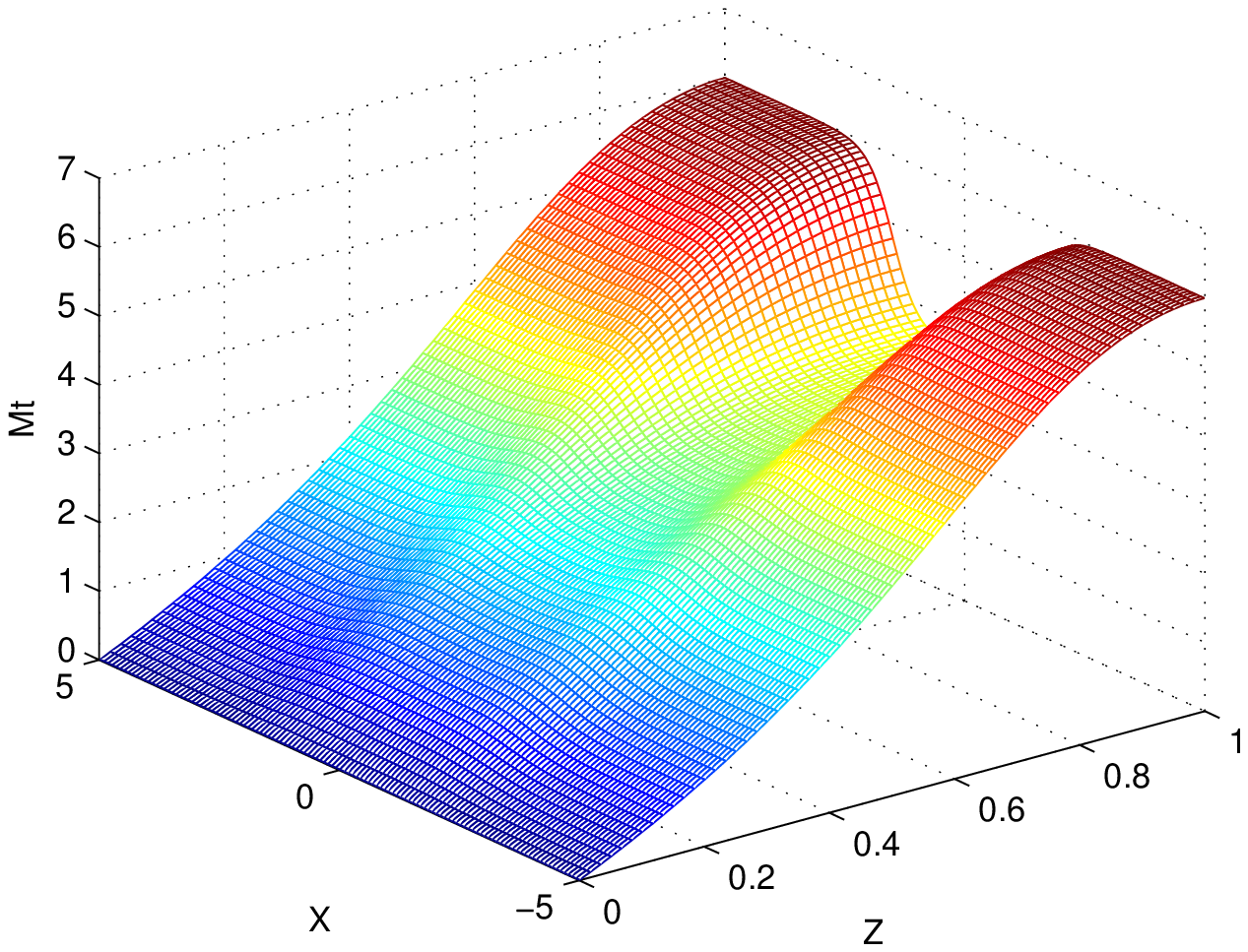}}
\hspace{1.0cm}
 \subfigure[$M_x$]{
  \includegraphics[width=0.4\textwidth]{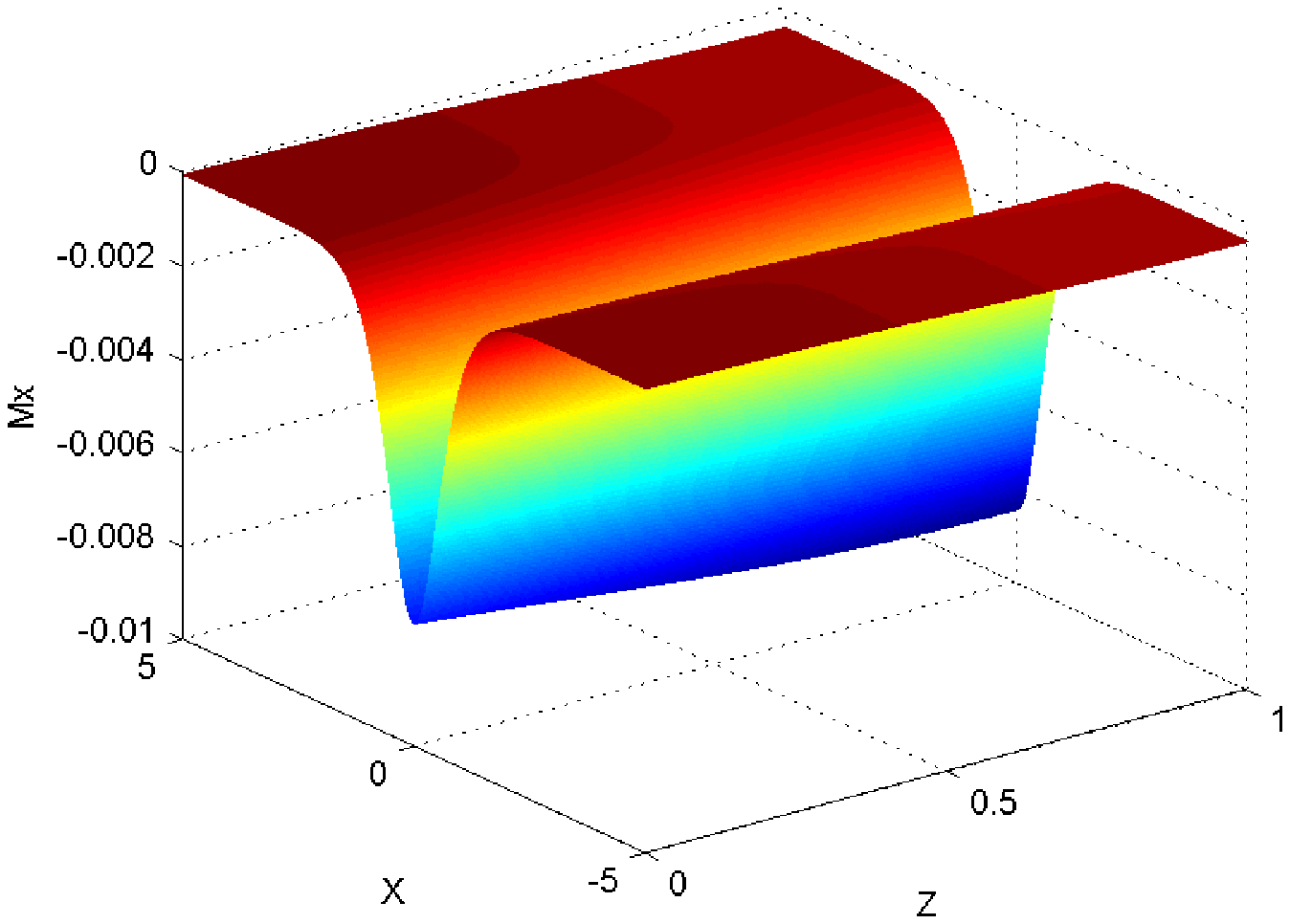}}
\end{center}  \vskip -5mm
\caption{Maxwell fields  $M_t$ and  $M_x$. The parameters are
set to $J/T^3_c=0.0103$, $\mu_\infty=6$, $L=3$, $\epsilon=0.6$, $\sigma=0.5$.}
 \label{fig_scalar_warpfactor}
\end{figure*}

In this section, we will computer the solution of the coupled equations (\ref{EO1})-(\ref{Mx}) numerically.  It is convenient to set the change of variables
$z=1-r_H/r$ and $\tilde x=\tanh(\frac{x}{4\sigma})$. First, we show the solutions for $M_t$ and $M_x$ in Fig. \ref{fig_scalar_warpfactor}, with $\mu_\infty=6$, $L=3$, $\epsilon=0.6$, and $\sigma=0.5$.

\begin{figure}
  \begin{center}
  \includegraphics[width=0.7\textwidth]{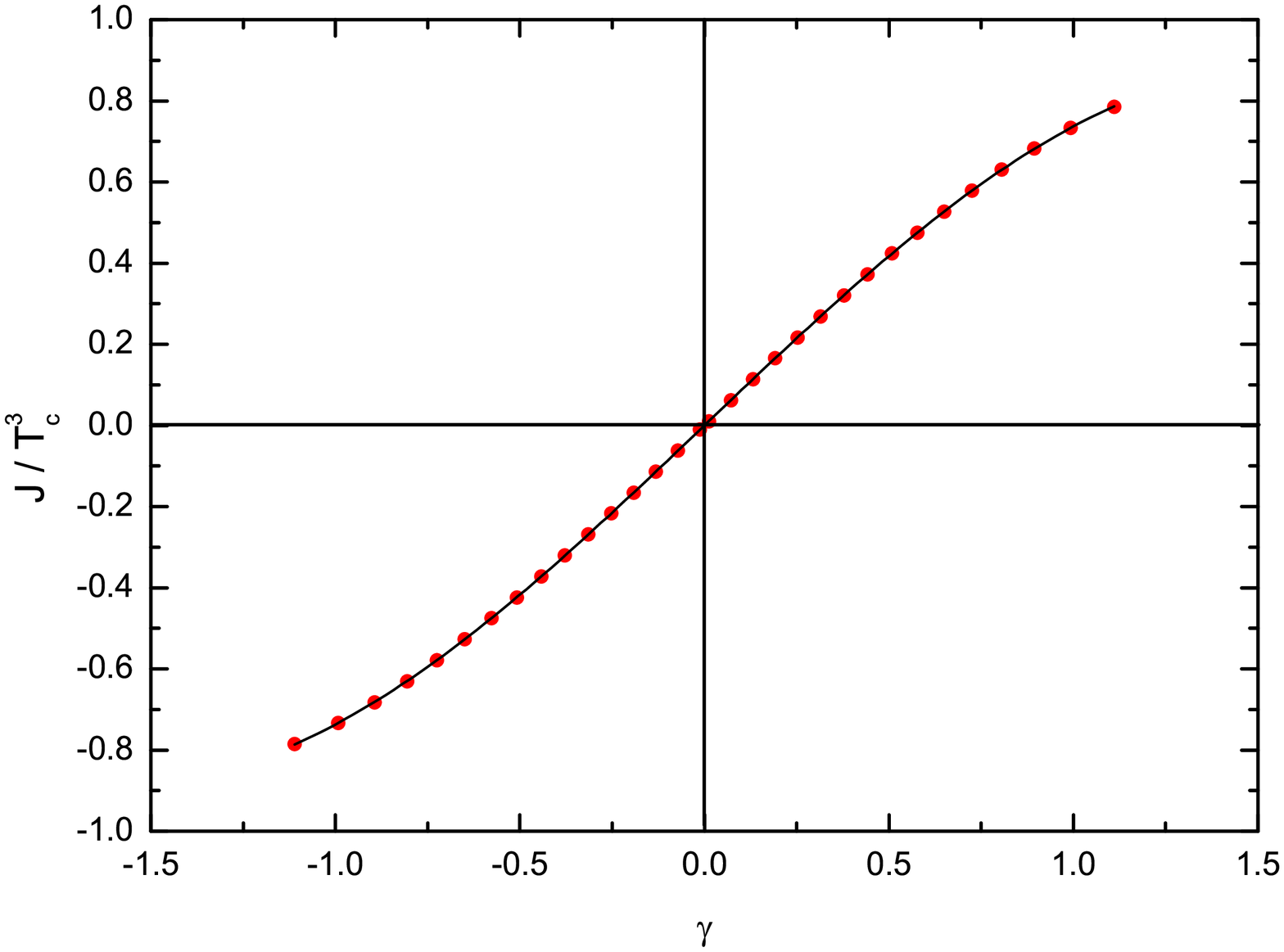}
  \end{center}
  \caption{Superfluid current $J_{\max}/T^3_c$ as the function of the phase difference $\gamma$. The black line is  the sine curve. The parameters are
set to $\mu_\infty=6$, $L=3$, $\epsilon=0.6$, $\sigma=0.5$.}\label{sine}
\end{figure}

Then, in Fig.~\ref{sine}, we find that the superfluid current is proportional to the sine of the phase difference across the junction, namely, the red dots come from the numerical calculations match precisely with the the black solid sine curve. Analyzing the graph, we can obtain the maximum current across the junction:
 \be
 J_{\max}/T^3_c \approx 0.871.
 \ee

\begin{figure*}
  \begin{center}
  \subfigure[$$]{{\label{figphi}}
  \includegraphics[width=0.7\textwidth]{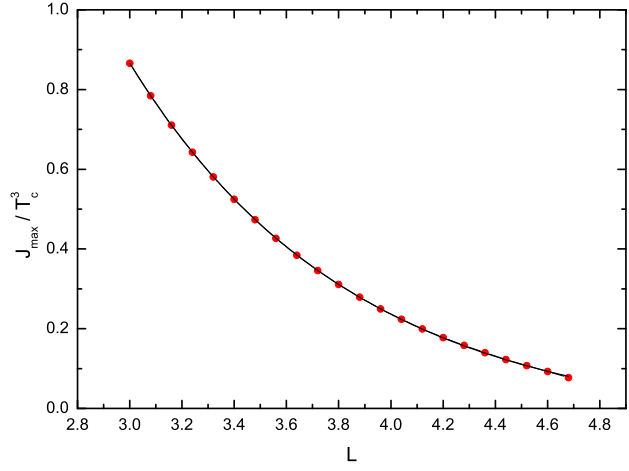}}
 \subfigure[$$]{{\label{figphiprime}}
  \includegraphics[width=0.7\textwidth]{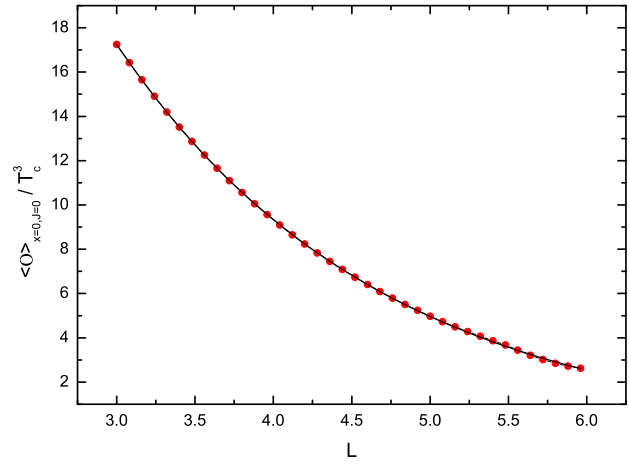}}
  \end{center}
\caption{The dependence of $J_{\max}$  and  $\langle\mathcal{O}\rangle_{x=0}$  on $L$. The parameters are
set to $\mu_\infty=6$, $\epsilon=0.6$, $\sigma=0.5$.}
 \label{fig2}
\end{figure*}

In Fig.~\ref{figphi}, the dependence of $J_{\max}$  on the width of the gap is shown. The graph predicts an exponential decay with the growing width of the gap in $J_{\max}$:
\be
J_{\max}/T^3_c = A_0\,e^{-\frac{\ell}{\xi}}.\label{jmaxell}
\ee
In Fig.~\ref{figphiprime}, the condensate $\langle\mathcal{O}\rangle=\psi^{(2)}$  at zero current is shown. The graph also predicts an exponential decay with the growing width of the gap in $J_{\max}$:
\be
\qquad \langle \mathcal{O}\rangle_{x = 0,J=0}/T^3_c= A_1\,e^{-\frac{\ell}{2\,\xi}}.\label{oell}
\ee
Fitting Eq.~(\ref{jmaxell}) and Eq.~(\ref{oell}) with the two sets of data, we can obtain $\{\xi ,A_0 \}  \approx \{0.85,30.32\}$  and $\{\xi,A_1\} \approx \{0.84,104.33\}$ for Eq.~(\ref{jmaxell}) and Eq.~(\ref{oell}), respectively.  The disagreement of two values of $\xi$ is smaller than the 4D case in \cite{Horowitz:2011dz}.

\begin{figure}
  \begin{center}
  \includegraphics[width=0.7\textwidth]{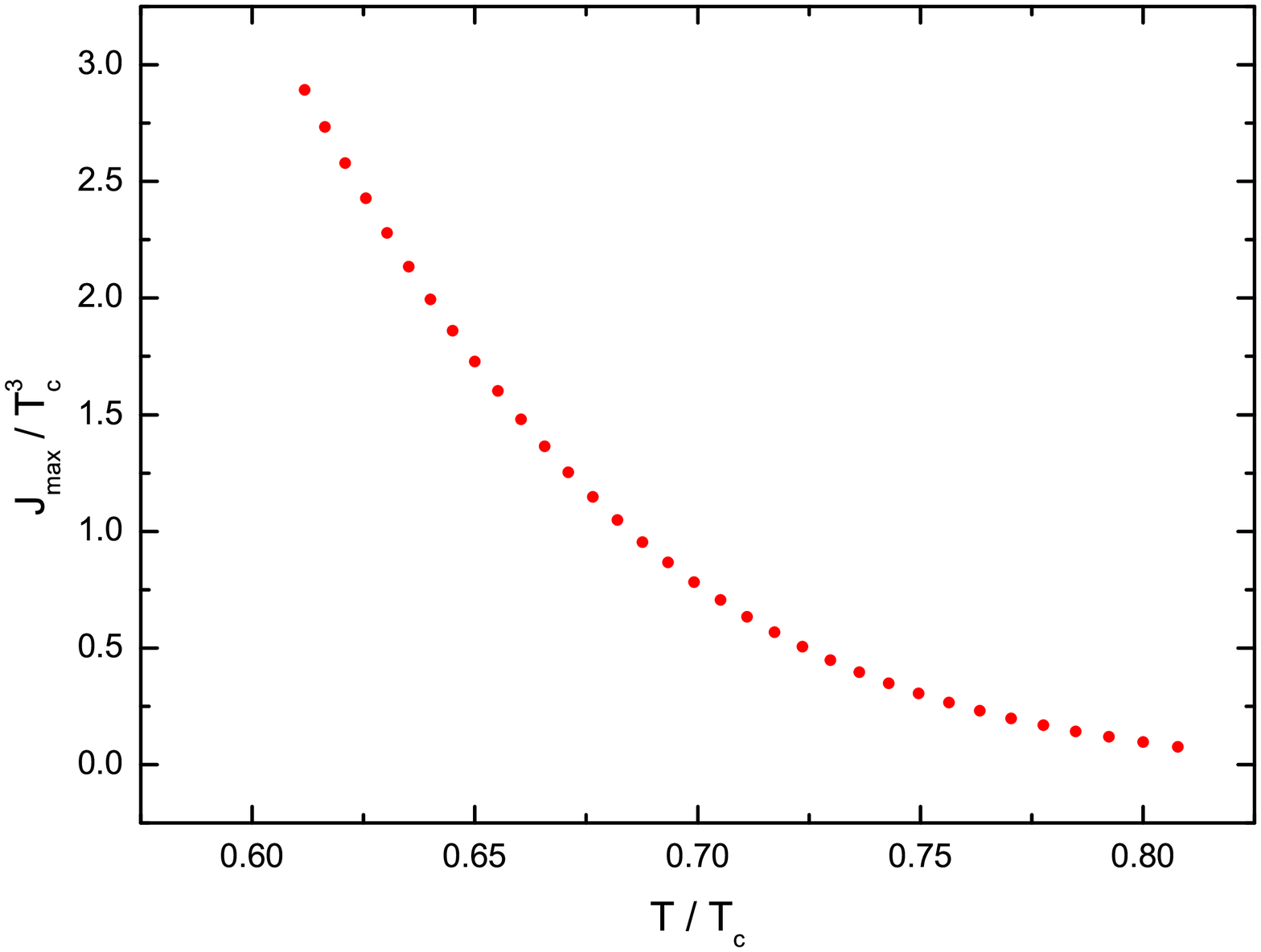}
  \end{center}
  \caption{The dependence of $J_{\max}$ on $T$, The parameters are
set to $\epsilon=0.6$, $\sigma=0.5$}\label{ffff}
\end{figure}

In Fig.~\ref{ffff}, we obtain the relation of $J_{\max}$ and $T$. Since $\epsilon=0.6$, we show the region corresponds to $T/T_c<0.6$, which depicts the character of an SNS Josephson Junction.

\section{Conclusion}

In this paper, the holographic description for the holographic DC Josephson junction
is extended to the case of the (3+1)-dimensional junction.  In the (4+1)-dimensional gravity background, we investigate a set of five couple, partial differential equations.  By choosing spatial dependence $\mu$ and non-vanishing constant $J$, we find the the superfluid current is proportional to the sine of the phase difference across the junction.
Moreover, the graph which predicts an exponential decay with the growing width of gap in $J_{\max}$ is obtained. At last,
the relation of $J_{\max}$ and $T$ is also obtained, from which one can verify that, near the critical temperature $T_c$, $J_{\max}$ can reach zero.

From the result of this paper, one can see that the model
of the holographic DC Josephson junction is valid for the (3+1)-dimensional case. It would be interesting to
investigate the holographic DC Josephson junction in
diverse dimensions and in various gravity theories.

\section*{Acknowledgement}

We would like to thank Lian-Chun YU for useful discussions.
This work was supported by the National Natural Science Foundation of China (No. 11005054 and No. 11075065), and the
Fundamental Research Fund for Physics and Mathematics of Lanzhou University (No.
LZULL200912).  Z.H. Zhao was supported by the Scholarship Award for Excellent Doctoral Student granted by Ministry of Education.


\begin{thebibliography}{99}

\bibitem{Maldacena:1997re}
  J.~M.~Maldacena,
  {\em The Large $N$ limit of superconformal field theories and supergravity},
  Adv.\ Theor.\ Math.\ Phys.\  {\bf 2}, 231 (1998)
  [Int.\ J.\ Theor.\ Phys.\  {\bf 38}, 1113 (1999)]
  [arXiv:hep-th/9711200].

\bibitem{Gubser:2008px}
  S.~S.~Gubser,
  {\em Breaking an Abelian gauge symmetry near a black hole horizon},
  Phys.\ Rev.\  {\bf D 78}, 065034 (2008)
  [arXiv:0801.2977 [hep-th]].

\bibitem{Hartnoll:2008vx}
  S.~A.~Hartnoll, C.~P.~Herzog and G.~T.~Horowitz,
  {\em Building a Holographic Superconductor},
  Phys.\ Rev.\ Lett.\  {\bf 101}, 031601 (2008)
  [arXiv:0803.3295 [hep-th]].

\bibitem{Hartnoll:2008kx}
  S.~A.~Hartnoll, C.~P.~Herzog and G.~T.~Horowitz,
  {\em Holographic Superconductors},
  JHEP {\bf 0812}, 015 (2008)
  [arXiv:0810.1563 [hep-th]].


\bibitem{Hartnoll:2009sz}
  S.~A.~Hartnoll,
  {\em Lectures on holographic methods for condensed matter physics},
  Class.\ Quant.\ Grav.\  {\bf 26}, 224002 (2009)
  [arXiv:0903.3246 [hep-th]].



\bibitem{Herzog:2009xv}
  C.~P.~Herzog,
  {\em Lectures on Holographic Superfluidity and Superconductivity},
  J.\ Phys.\ {\bf A  42}, 343001 (2009)
  [arXiv:0904.1975 [hep-th]].

\bibitem{Horowitz:2010gk}
  G.~T.~Horowitz,
  {\em Introduction to Holographic Superconductors},
  arXiv:1002.1722 [hep-th].


\bibitem{Horowitz:2011dz}
  G.~T.~Horowitz, J.~E.~Santos and B.~Way,
  {\em A Holographic Josephson Junction},
  arXiv:1101.3326 [hep-th].


\bibitem{Basu:2008st}
  P.~Basu, A.~Mukherjee, H.~-H.~Shieh,
  {\em Supercurrent: Vector Hair for an AdS Black Hole},
  Phys.\ Rev.\  {\bf D 79}, 045010 (2009)
  [arXiv:0809.4494 [hep-th]].

\bibitem{Herzog:2008he}
  C.~P.~Herzog, P.~K.~Kovtun, D.~T.~Son,
  {\em Holographic model of superfluidity},
  Phys.\ Rev.\  {\bf D 79}, 066002 (2009)
  [arXiv:0809.4870 [hep-th]].

\bibitem{Arean:2010xd}
  D.~Arean, M.~Bertolini, J.~Evslin and T.~Prochazka,
  {\em On Holographic Superconductors with DC Current},
  JHEP {\bf 1007}, 060 (2010)
  [arXiv:1003.5661 [hep-th]].

\bibitem{Sonner:2010yx}
  J.~Sonner, B.~Withers,
  {\em A gravity derivation of the Tisza-Landau Model in AdS/CFT},
  Phys.\ Rev.\  {\bf D 82}, 026001 (2010)
  [arXiv:1004.2707 [hep-th]].





\bibitem{Horowitz:2008bn}
  G.~T.~Horowitz and M.~M.~Roberts,
  {\em Holographic Superconductors with Various Condensates},
  Phys.\ Rev.\  {\bf D 78}, 126008 (2008)
  [arXiv:0810.1077 [hep-th]].




\bibitem{Breitenlohner:1982bm}
  P.~Breitenlohner and D.~Z.~Freedman,
  {\em Positive Energy in anti-De Sitter Backgrounds and Gauged Extended
  Supergravity},
  Phys.\ Lett.\  {\bf B 115}, 197 (1982).



\bibitem{Arean:2010zw}
  D.~Arean, P.~Basu and C.~Krishnan,
  {\em The Many Phases of Holographic Superfluids},
  JHEP {\bf 1010}, 006 (2010)
  [arXiv:1006.5165 [hep-th]].












\end{thebibliography}
\end{document}